\documentclass[10pt]{llncs}
\usepackage[english]{babel}
\usepackage[inline]{enumitem}
\setlist[enumerate,1]{label=\textit{\roman*)}}
\usepackage{listings}
\usepackage{subfig}
\usepackage{graphicx}
\usepackage{amsmath}
\usepackage{color}
\usepackage{xspace}
\usepackage{multirow}

\usepackage{tikz}
 \usetikzlibrary{fit}
\usetikzlibrary{matrix}
\usetikzlibrary{positioning}
\usetikzlibrary{shapes,snakes}
\usetikzlibrary{arrows}
\usepackage{tikz}  
\usetikzlibrary{fit,positioning,hobby}
\usetikzlibrary{calc}
\usetikzlibrary{decorations.text}
\usetikzlibrary{shapes}

\usepackage{cite}

\usetikzlibrary{trees}

\usetikzlibrary{decorations.markings,matrix}
\usetikzlibrary{decorations.text,calc}

\usepackage{algpseudocode}
\algrenewcommand\textproc{\textit}
\usepackage{algorithm}

\newcommand{\opti}{\emph{Odyssey}\xspace}
\newcommand{\hibiscus}{HiBISCuS}

\newcommand{\remove}[1]{{}}

\usepackage{times}

\usepackage{enumitem}
\setitemize{noitemsep,topsep=0pt,parsep=0pt,partopsep=5pt}
\setenumerate{noitemsep,topsep=0pt,parsep=0pt,partopsep=2pt}
\setdescription{noitemsep,topsep=0pt,parsep=0pt,partopsep=4pt}

\makeatletter
\renewcommand\section{\@startsection{section}{1}{\z@}
                     {-10\p@ \@plus -4\p@ \@minus -4\p@}
                     {3\p@ \@plus 2\p@ \@minus 2\p@}
                     {\normalfont\large\bfseries\boldmath
                      \rightskip=\z@ \@plus 8em\pretolerance=10000 }}
\renewcommand\subsection{\@startsection{subsection}{2}{\z@}
                      {-6\p@ \@plus -4\p@ \@minus -4\p@}
                      {2\p@ \@plus 4\p@ \@minus 4\p@}
                      {\normalfont\normalsize\bfseries\boldmath
                       \rightskip=\z@ \@plus 8em\pretolerance=10000 }}
\renewcommand\subsubsection{\@startsection{subsubsection}{3}{\z@}
                     {-3\p@ \@plus -4\p@ \@minus -4\p@}
                     {-0.5em \@plus -0.22em \@minus -0.1em}
                     {\normalfont\normalsize\bfseries\boldmath}}
\renewcommand\paragraph{\@startsection{paragraph}{4}{\z@}
                      {-4\p@ \@plus -4\p@ \@minus -4\p@}
                      {-0.5em \@plus -0.22em \@minus -0.1em}
                      {\normalfont\normalsize\bfseries\boldmath}}     

\makeatother

\begin{document}

\title{The \opti Approach for Optimizing Federated SPARQL Queries}

\author{Gabriela Montoya\inst{1} \and Hala Skaf-Molli\inst{2} \and Katja Hose\inst{1}}

\institute{Aalborg University, Denmark
    \email{\{gmontoya,khose\}@cs.aau.dk} \and
    Nantes University, France
    \email{hala.skaf@univ-nantes.fr}}
    
\maketitle
    
\begin{abstract}
Answering queries over a federation of SPARQL endpoints requires combining data from more than one data source. Optimizing queries in such scenarios is particularly challenging not only because of (i) the large variety of possible query execution plans that correctly answer the query but also because (ii) there is only limited access to statistics about schema and instance data of remote sources. To overcome these challenges, most federated query engines rely on heuristics to reduce the space of possible query execution plans or on dynamic programming strategies to produce optimal plans. Nevertheless, these plans may still exhibit a high number of intermediate results or high execution times because of heuristics and inaccurate cost estimations. In this paper, we present \opti, an approach that uses statistics that allow for a more accurate cost estimation for federated queries and therefore enables \opti to produce better query execution plans. Our experimental results show that  \opti produces query execution plans that are better in terms of data transfer and execution time than state-of-the-art optimizers. Our experiments using the FedBench benchmark show execution time gains of at least 25 times on average.
\end{abstract}
    
\section{Introduction}

Federated SPARQL query engines~\cite{DBLP:conf/semweb/AcostaVLCR11,DBLP:journals/ws/BascaB14,DBLP:conf/semweb/GorlitzS11,DBLP:conf/esws/QuilitzL08,DBLP:conf/semweb/SchwarteHHSS11} answer SPARQL queries over a federation of SPARQL endpoints. Query optimization is a particularly complex and challenging task in a federated setting. %~\cite{DBLP:series/sci/GorlitzS11}
The query optimizer minimizes processing and communication costs by selecting only relevant sources for a query. It decomposes the query into subqueries, and produces a query execution plan with good join ordering and physical operators. With limited access to statistics, however, most federated query engines rely on heuristics~\cite{DBLP:conf/semweb/AcostaVLCR11,DBLP:conf/semweb/SchwarteHHSS11} to reduce the huge space of possible plans or on dynamic programming (DP)~\cite{DBLP:conf/i-semantics/CharalambidisTK15,DBLP:conf/semweb/GorlitzS11} to produce optimal plans. However, these plans may still exhibit a high number of intermediate results or high execution times because of inadequate heuristics or inaccurate estimations of cost functions~\cite{DBLP:conf/edbt/Gubichev014}.

In this paper,  we propose \opti, a cost-based query optimization approach for federations of SPARQL endpoints. 
\opti defines statistics for representing entities inspired by \cite{DBLP:conf/icde/NeumannM11} and  statistics for representing links among datasets while guaranteeing result completeness.
   In a federated setting, computing statistics naturally requires access to more than one dataset. 
To reduce the overhead, \opti uses entity synopsis to identify links among datasets.
This comes at the risk of losing some accuracy in the link identification but still guarantees that no links will be missed during query optimization, i.e., there is a small risk that more sources are queried than strictly necessary but the query result will be complete. 

\opti uses the computed statistics to estimate the sizes of intermediate results and dynamic programming to produce an efficient query execution plan with a low number of intermediate results. 
In summary, this paper makes the following contributions:
\begin{itemize}[label={$\bullet$}]
    \item Concise statistics of adequate granularity representing entities and describing links among datasets while guaranteeing result completeness.
    \item A lightweight technique to compute federated statistics in a federated setup that relies on entity synopsis. 
    \item A query optimization algorithm based on dynamic programming using our statistics to find the best plan. 
    \item Extensive evaluation using a well-accepted standard benchmark for federated query processing~\cite{DBLP:conf/semweb/SchmidtGHLST11}, including comparison against a broad range of state-of-the-art related work~\cite{DBLP:conf/i-semantics/CharalambidisTK15,DBLP:conf/semweb/GorlitzS11,DBLP:conf/esws/SaleemN14,DBLP:conf/semweb/SchwarteHHSS11}. The results show \opti's superiority with a speed-up of up to 126 times and a reduction of transferred data of up to 118 times on average.
\end{itemize}
This paper is organized as follows.
Section~\ref{sec:related-work} presents related work, 
Section~\ref{sec:proposal} describes the \opti approach and its  algorithms.
Section~\ref{sec:experimental-results} discusses our experimental results.
Finally, conclusions and future work are outlined in Section~\ref{sec:conclusion}.

\section{Related Work}
\label{sec:related-work}

Query optimization in state-of-the-art federated query engines, such as FedX~\cite{DBLP:conf/semweb/SchwarteHHSS11} and ANAPSID~\cite{DBLP:conf/semweb/AcostaVLCR11}, 
relies on heuristics. For instance, FedX~\cite{DBLP:conf/semweb/SchwarteHHSS11} integrates the variable counting heuristic, %~\cite{DBLP:conf/www/StockerSBKR08}, 
where 
relative selectivity of triple patterns is heuristically estimated according to the presence of constants and variables in the triple patterns.
These heuristics are lightweight but might not lead to the best query execution plan~\cite{DBLP:conf/www/StockerSBKR08}. 
To find an optimal plan, several approaches~\cite{DBLP:conf/i-semantics/CharalambidisTK15,DBLP:conf/semweb/GorlitzS11,DBLP:conf/esws/QuilitzL08,DBLP:conf/www/WangTD13} rely on dynamic programming.
However, given the high number of alternative query plans for SPARQL queries with many triple patterns,  dynamic programming is very expensive~\cite{DBLP:conf/edbt/Gubichev014}. 
Another important factor of query optimization is source selection. Several approaches~\cite{DBLP:conf/semweb/SchwarteHHSS11,DBLP:conf/semweb/AcostaVLCR11,DBLP:conf/semweb/GorlitzS11,DBLP:conf/esws/SaleemN14,DBLP:conf/www/WangTD13} try to determine the relevance of a source by sending ASK queries, which increases the costs for a single query but might amortize in large federations for an overlapping query load. 
Another technique is to estimate whether combining the data of multiple sources can lead to any join results, e.g., by computing the intersection of the sources' URI authorities~\cite{DBLP:conf/esws/SaleemN14} or detailed statistics~\cite{DBLP:conf/www/HarthHKPSU10,DBLP:conf/edbt/PrasserKK12}.
\remove{In Section~\ref{sec:experimental-results}, we show that even if the optimization time of \opti can be slightly longer to the above mentioned approaches, \opti obtains plans that are executed faster than the other approaches.}

Federated query optimization can also rely on cardinality estimations based on statistics and used, for instance, to reduce sizes of intermediate results.
Most available statistics~\cite{DBLP:conf/semweb/ArandaHUV13} use the Vocabulary of Interlinked Datasets voiD~\cite{DBLP:conf/www/AlexanderCHZ09}, which describes statistics at dataset level (e.g., the number of triples), at the property level (e.g., for each property, its number of different subjects), and at the class level (e.g., the number of instances of each class).
However, approaches based on voiD~\cite{DBLP:conf/i-semantics/CharalambidisTK15,DBLP:conf/semweb/HagedornHSU14,DBLP:conf/semweb/GorlitzS11} and other statistics, such as QTrees~\cite{DBLP:conf/www/HarthHKPSU10} and PARTrees~\cite{DBLP:conf/edbt/PrasserKK12}, share the drawback of missing the best query execution plans because of errors in estimating cardinalities caused by relying on assumptions that often do not hold for arbitrary RDF datasets~\cite{DBLP:conf/icde/NeumannM11}, e.g., a uniform data distribution and that the results of triple patterns are independent. 

Characteristic sets (CS)~\cite{DBLP:conf/dasfaa/DuCD12,DBLP:conf/icde/NeumannM11} aim at solving this problem in centralized systems by capturing statistics about sets of entities having the same set of properties. This information can then be used to accurately estimate the cardinality and join ordering of star-shaped queries. Typically, any set of joined triple patterns in a query can be divided into connected star-shaped subqueries. Subqueries in combination with the predicate that links them, define a characteristic pair (CP)~\cite{DBLP:conf/edbt/Gubichev014,icde/Meimaris17}. Statistics about such CPs can then be used to estimate the selectivity of two star-shaped subqueries. 
Such cardinality estimations can be combined with dynamic programming on a reduced space of alternative query plans.  
Whereas existing work on CSs and CPs were developed for centralized environments, this paper proposes a solution generalizing these principles for federated environments.

\section{The \opti Approach}
\label{sec:proposal}
Inspired by the latest advances in statistics for centralized triple stores~\cite{DBLP:conf/icde/NeumannM11,DBLP:conf/edbt/Gubichev014,icde/Meimaris17}, \opti uses statistics about individual datasets to derive detailed statistics for optimizing federated queries.
In the following, we first describe the foundations of our statistics on individual datasets (Section~\ref{sec:cp}) and then propose a novel method for computing such statistics in a federated environment based on entity descriptions (Section~\ref{sec:fedStats}). 
As the detailed entity descriptions cause too much overhead in a federated setup, we propose a method for reducing the sizes of the descriptions (Section~\ref{sec:reduceSizes}). 
Finally, we present the \opti approach for query optimization and its main steps (Section~\ref{sec:optimization}): source selection, join ordering, and query decomposition.

\subsection{Dataset Statistics on Individual Datasets}
\label{sec:cp}

\subsubsection{Star-Shaped Subqueries}

To estimate the cardinality and costs of BGPs sharing the same subject (or object), i.e., \textit{star-shaped subqueries}, we exploit the principle that entities sharing the same set of properties are similar. 
In this context, we refer to the set of an entity's properties as its characteristic set (CS) and use \textit{cs$_s$(e)} to denote the CS of entity \textit{e} in dataset \textit{s} or \textit{cs(e)} if \textit{s} is clear from the context. 
For instance, in DBpedia 3.5.1  \textit{cs(dbr:Gary\_Goetzman)}=$C_1$=\{\textit{dbo:birthDate}, \textit{foaf:name}, \textit{rdf:type}, \textit{dbo:activeYearsStartYear}, \textit{rdfs:label}, \textit{skos:subject}\}. In total, 260 entities share this set of properties and therefore CS $C_1$. 

\begin{lstlisting}[basicstyle=\scriptsize\sffamily,caption={Statistics for CS $C_1$},label=lst:statsCS,numbers=none,frame=none,columns=fixed,extendedchars=true,breaklines=true,showstringspaces=false]
{ count: 260,
  elems:
  {{ pred: dbo:birthDate, ocurrences : 260 },
   { pred: foaf:name, ocurrences: 326 },
   { pred: rdf:type, ocurrences: 1023 },
   { pred: dbo:activeYearsStartYear, ocurrences: 260 },
   { pred: rdfs:label, ocurrences: 260 },
   { pred: skos:subject, ocurrences: 1336 }}}
\end{lstlisting}

CSs can be computed by scanning once a dataset's triples sorted by subject; after all the entity properties have been scanned, the entity's CS is identified. 
For each CS \textit{C}, we compute statistics, i.e., the number of entities sharing \textit{C} (\textit{count(C)}) and the number of triples with predicate \textit{p} occurring with these entities 
(\textit{occurrences(p, C)}). 
Listing~\ref{lst:statsCS} shows the statistics for the above mentioned example CS $C_1$. Entities of $C_1$ occur on average in 1 triple with property \textit{dbo:birthDate} and in 3.94 triples with property \textit{rdf:type}. 

For a star-shaped query, only CSs including all of the query's properties are relevant as entities that only satisfy a subset of these properties cannot contribute to the answer.

\begin{lstlisting}[basicstyle=\scriptsize\sffamily,caption={Find persons that have been active},label=queryB,language=sparql,numbers=none,frame=none,columns=fixed,extendedchars=true,breaklines=true,showstringspaces=false]
SELECT DISTINCT ?person WHERE {
  ?person dbo:birthDate ?date .          (tp1)
  ?person dbo:activeYearsStartYear ?sy . (tp2)
  ?person foaf:name ?name                (tp3)
}
\end{lstlisting}
For star-shaped queries asking for the set of unique entities described by some properties (query with DISTINCT modifier), the exact number of answers can be determined precisely (no estimation). For example, the cardinality of the query given in Listing~\ref{queryB} can be obtained by adding up the \textit{count(C)} of all \textit{CS}s containing the properties \textit{dbo:birthDate}, \textit{dbo:activeYearsStartYear}, and \textit{foaf:name}.
In DBpedia 3.5.1, there are 7,059 CSs that include these three properties, and the total number of entities with these CSs is 83,438.
Formally, the number of entities for a given set of properties \textit{P}, \textit{cardinality(P)}, is computed based on the CSs \textit{C$_j$} that include all the properties in \textit{P} as:

\begin{equation} \label{f:cost} \scriptsize
cardinality(P) = \sum_{P \subseteq C_j} count(C_j)
\end{equation}  
For queries without the DISTINCT modifier, we need to account for duplicates by considering the number of triples with predicate $p_i \in P$ that an entity is associated with on average: 

\begin{equation} \label{f:estimationMult} \scriptsize
\hspace*{-0.5cm}
estimatedCardinality(P) = \sum_{P \subseteq C_j} \Big( count(C_j) * \prod_{p_i \in P} \frac{ocurrences(p_i, C_j)}{count(C_j)}\Big)
\end{equation}
In DBpedia 3.5.1, as mentioned above, there are 7,059 CSs relevant for the query in Listing~\ref{queryB} with 83,438 entities as answer. These 83,438 entities are described by  109,830 triples with predicate \textit{foaf:name}, 83,448 with predicate \textit{dbo:birthDate}, and 110,460 with predicate \textit{dbo:activeYearsStartYear}. If the query is considered without the DISTINCT modifier, i.e., considering duplicated results, we estimate: 148,486  
matching entities in the result, which is very close to the real number (149,440).

Once the relevant CSs for a query have been identified, they can be used to find the join order minimizing the sizes of intermediate results.
For the query in Listing~\ref{queryB}, we start by estimating the cardinalities for each subquery with two out of the three triple patterns using Formula~\ref{f:cost}: \{tp1, tp2\}: 98,281, \{tp1, tp3\}: 209,731, and \{tp2, tp3\}: 127,712. 
The triple pattern not included in the cheapest subquery (\{tp1, tp2\}) is executed last (tp3). 
We proceed recursively with the cheapest subquery and determine the cardinalities for its subsets: \{tp1\}: 232,608 and \{tp2\}: 143,004. Again, the triple pattern not included in the cheapest subquery (tp1) will be executed last of the currently considered set of triple patterns.
As a result, we will execute the join between tp2 and tp1 first and afterwards compute the join with tp3. We also get the order in which the triple patterns should be evaluated for the first join: first tp2 and then tp1.

\subsubsection{Arbitrary Queries}

To estimate the cardinality for queries with more complex shapes, we need to consider the connections (links) between entities with different CSs. Entity \textit{dbr:Evan\_Almighty}, for example, is linked to \textit{dbr:Tom\_Shadyac} via property  \textit{dbo:director} by triple \textit{(dbr:Evan\_Almighty, dbo:director, dbr:Tom\_Shadyac)}. 

The links between CSs via properties can formally be described by characteristic pairs (CPs), they are 
 defined as \textit{(cs$_{s}$(e1), cs$_{s}$(e2), p)} for entities e1 and e2 if \textit{(e1, p, e2)} $\in$ \textit{s}. 
The statistics -- \textit{$count((C_i,C_j,p))$} -- capture the number of links between a pair of CSs ($C_i$ and $C_j$) using a particular property $p$. 
For example, given the CSs of \textit{dbr:Evan\_Almighty} and \textit{dbr:Tom\_Shadyac} as $C_1$ and $C_2$ the number of links via property \textit{dbo:director} is given by: \textit{count}($(C_1$, $C_2$, \textit{dbo:director})). 

\begin{lstlisting}[basicstyle=\scriptsize\sffamily,caption={Find movies and their directors},label=queryC,language=sparql,numbers=none,frame=none,columns=fixed,extendedchars=true,breaklines=true,showstringspaces=false]
SELECT DISTINCT ?film ?director WHERE {
  ?film dbo:runtime ?runtime .             (tp1)
  ?film dbo:director ?director .           (tp2)
  ?film dbo:budget ?budget .               (tp3)
  ?director dbo:birthDate ?date .          (tp4)
  ?director dbo:activeYearsStartYear ?sy . (tp5)
  ?director foaf:name ?name                (tp6)
}
\end{lstlisting}
The number of unique results (pairs of entities with set of properties $P_k$ and $P_l$, query with DISTINCT modifier) can be exactly computed (not estimated) using the formula: 
\begin{equation} \label{f:estimationMultCP} \scriptsize
\begin{split}
cardinality((P_k, P_l, p)) & = \sum_{P_k \subseteq C_i \land P_l \subseteq C_j} count((C_i, C_j, p))
\end{split}
\end{equation}
For the query in Listing~\ref{queryC} property \textit{dbo:director} links several pairs of CSs representing movies and actors. Hence, we need to compute  $\Sigma_{f_1 \land f_2}$ $count$(($C_i$, $C_j$, \textit{dbo:director})), where $f_1$ is (\{\textit{dbo:runtime}, \textit{dbo:director}, \textit{dbo:budget}\} $\subseteq$ $C_i$) and $f_2$ is (\{\textit{dbo:birthDate}, \textit{dbo:activeYearsStartYear}, \textit{foaf:name}\} $\subseteq$ $C_j$); one of the operands of this sum is $count$(($C_1$, $C_2$, \textit{dbo:director})) mentioned in the example above.
For this query, DBpedia 3.5.1 contains 1,509 CPs linking entities from two CSs via property \textit{dbo:director}.   

If a query does not involve the DISTINCT modifier, result cardinality estimation considers the property occurrences in the CSs: 

\begin{equation} \label{f:estimationMultCP} \scriptsize
\begin{split}
\hspace*{-0.25cm}
estimatedCardinality((P_k, P_l, p)) & = \sum_{P_k \subseteq C_i \land P_l \subseteq C_j} \Big(count((C_i, C_j, p))\\
&\hspace*{-2cm}*\prod_{p_k \in P_k-\{p\}}\big(\frac{ocurrences(p_k, C_i)}{count(C_i)}\big)
                                  *\prod_{p_l \in P_l}\big(\frac{ocurrences(p_l, C_j)}{count(C_j)}\big)\Big)
\end{split}
\end{equation}
\remove{Cardinality estimation considers the average number of triples per query predicate in the relevant CPs. Selectivity of predicate \textit{p} is not considered in the product because it is already considered by $count(C_k,C_l,p)$.}
\noindent Assuming that the order of joins within star-shaped subqueries has already been optimized based on the CSs as described above, we treat each star-shaped subquery as a single meta-node to reduce complexity. We estimate the cardinalities of joins between the meta-nodes using the statistics on CPs and use dynamic programming (DP) to determine the optimal join order that minimizes the sizes of intermediate results.
Although the presentation in this section focuses on subject-subject joins, the same principle can be applied to other types of joins, e.g., object-object.

\begin{figure*}[htb]
\centering
\includegraphics[width=1\textwidth]{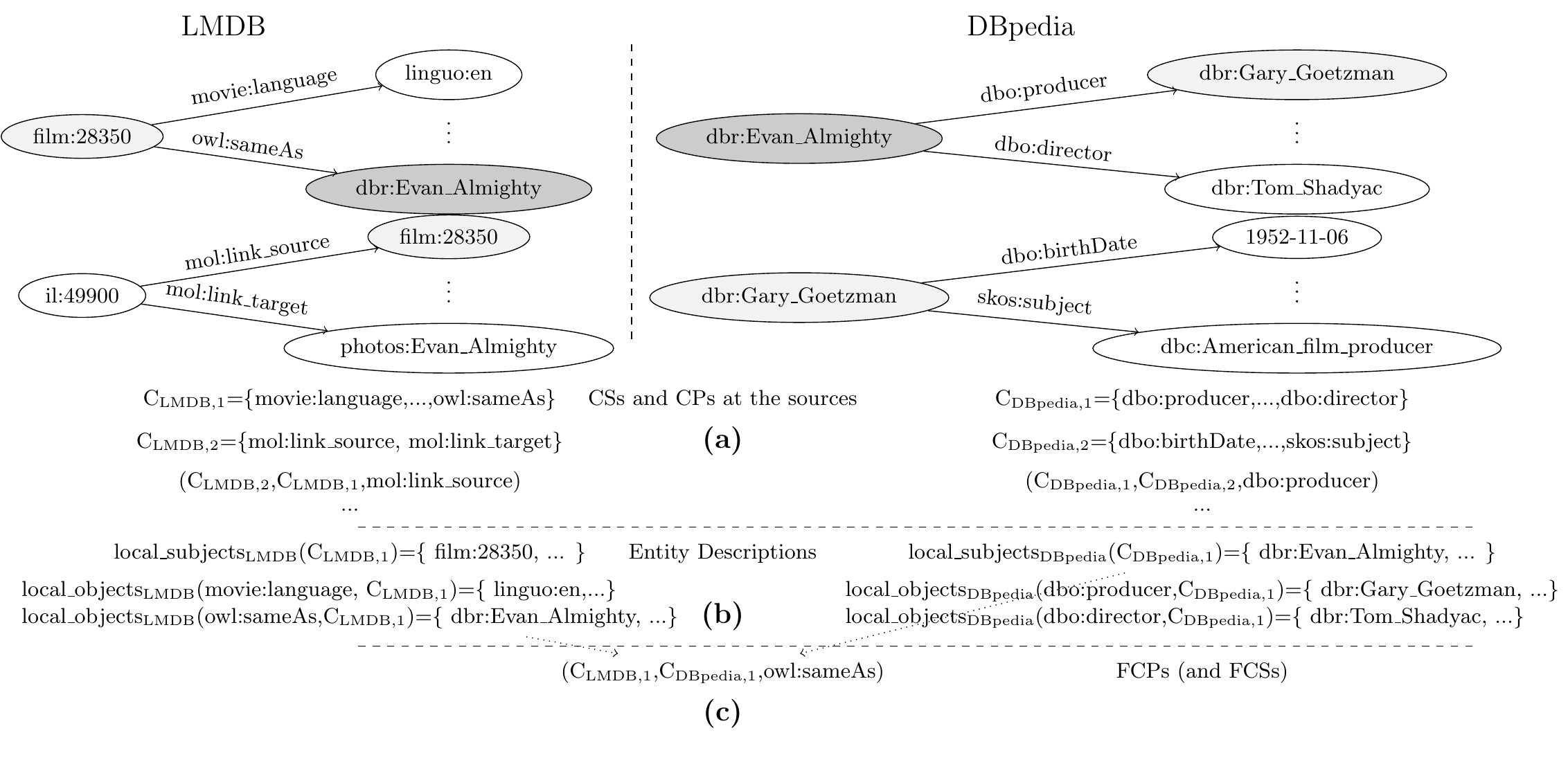}
\caption{Federated Computation of Statistics} 
\label{fig:federatedStatistics}
\end{figure*}

\subsection{Federated Statistics}
\label{sec:fedStats}

In general, entities might occur in multiple datasets in a federation $S$. Hence, we define a \textit{federated characteristic set} (FCS) as follows: 
\textit{fcs$_S$(e)= $\bigcup_{s \in S}$ cs$_s$(e)}, $S$  might be omitted if clear from the context. 
However, triples describing the same entity are typically part of a single dataset so that most CSs can be computed over each dataset independently from the others\footnote{FCSs describing entities across multiple datasets are very rare. In FedBench, for instance, they affect less than 0.5\% of all CSs.}. 
The \textit{federated characteristic pair} (FCP) of entities \textit{e1} and \textit{e2} 
via property \textit{p} in federation \textit{S} is defined as \textit{(fcs$_{S}$(e1), fcs$_{S}$(e2), p)}.
For FCSs FC$_i$ and FC$_j$ and property $p$, we compute statistics \textit{count(FC$_i$)}, \textit{occurrences(p, FC$_i$)}, and \textit{count((FC$_i$,FC$_j$,p))} as before for CSs and CPs. 
For simplicity, the following sections focus on FCPs connecting CSs instead of FCSs. The generalization using FCSs is straightforward.

Whereas single dataset statistics can be computed once and provided by the sources in the same way they currently provide voiD statistics~\cite{DBLP:conf/www/AlexanderCHZ09}, FCSs and FCPs require more effort and centralized knowledge about all entities in the considered datasets. 
A naive way to compute FCSs and FCPs is evaluating expensive SPARQL queries with FILTER expressions involving NOT EXISTS, but this can take weeks for a dataset with thousands of CSs. 
It is much more efficient if the sources directly share information about local subjects and objects with the federated query engine: \textit{local\_subjects$_s$}(C) contains the IRIs of entities with CS C for source s, while \textit{local\_objects$_s$}(p, C) contains the IRIs of entities linked via predicate p to subjects with CS C.
Such information can, for instance, be obtained efficiently while computing CSs and CPs locally and then shared with the federated query engine. 

The federated query engine can then use this information to compute FCSs and FCPs. Consider, for instance, 
the two datasets LMDB and DBpedia in Fig.~\ref{fig:federatedStatistics}; 
based on the CSs (Fig.~\ref{fig:federatedStatistics}(a)), the sources compute \textit{entity descriptions} (\textit{local\_subjects$_i$} and \textit{local\_objects$_i$} in Fig.~\ref{fig:federatedStatistics}(b)). 
Entity \textit{film:28350} has properties \{movie:language,...,owl:sameAs\}=$C_{\text{LMDB},1}$. Hence, \textit{film:28350} $\in$ \textit{local\_subjects$_{\text{LMDB}}$}($C_{\text{LMDB},1}$). There is a triple with \textit{dbr:Evan\_Almighty} as  value of property \textit{owl:sameAs} for an entity with CS $C_{\text{LMDB},1}$  (\textit{film:28350}) so \textit{dbr:Evan\_Almighty} $\in$ \textit{local\_objects$_{\text{LMDB}}$}(owl:sameAs, $C_{\text{LMDB},1}$) (Fig.~\ref{fig:federatedStatistics}(b)).
The overlap between the set of entities  \textit{local\_subjects$_{\text{DBpedia}}$}($C_{\text{DBpedia},1}$) and \textit{local\_objects$_{\text{LMDB}}$}(owl:sameAs, $C_{\text{LMDB},1}$) represent linked entities between LMDB and DBpedia via property owl:sameAs. 
Hence, we obtain FCP ($C_{\text{LMDB},1}$, $C_{\text{DBpedia},1}$, owl:sameAs) (Fig.~\ref{fig:federatedStatistics}(c)). \textit{count}(($C_{\text{LMDB},1}$, $C_{\text{DBpedia},1}$, owl:sameAs)) corresponds to the cardinality of the intersection between all the \textit{local\_objects$_{\text{DBpedia}}$} and \textit{local\_subjects$_{\text{LMDB}}$} linked by property owl:sameAs. 

\begin{algorithm*}[htb]
\scriptsize
\caption{Compute FCPs Algorithm}
\label{algo:computeCP}
\begin{algorithmic}[1]
\Statex \textbf{Input:} \textit{local\_objects$_{d1}$} and \textit{local\_subjects$_{d2}$} for datasets $d1$ and $d2$
\Statex \textbf{Output:} A set of FCPs (\textit{FCPs}) with links from $d1$ to $d2$; \textit{count}(fcp) for each fcp in FCPs
\Function{ComputeFCPs}{local\_subjects$_{d2}$, local\_objects$_{d1}$}
\State FCPs $\leftarrow$ \{ \}
\State count $\leftarrow$ newFunction(0)%newFunctionWithDefaultValue(0) %\{ \}
\For {(p, C$_{d1,i}$) $\in$ domain(local\_objects$_{d1}$)}
          \State entities $\leftarrow$ local\_objects$_{d1}$(p,C$_{d1,i}$)
          \For{C$_{d2,j}$ $\in$ domain(local\_subjects$_{d2}$)}
              \State entities $\leftarrow$ entities $\bigcap$ local\_subjects$_{d2}$(C$_{d2,j}$) ~\label{lst:line:overlap}
              \If {entities $\neq$ $\emptyset$}
                  \State FCPs $\leftarrow$ FCPs $\bigcup$ \{ (C$_{d1,i}$, C$_{d2,j}$, p) \} \label{lst:line:newcp}
                  \State count((C$_{d1,i}$, C$_{d2,j}$, p)) $\leftarrow$ count((C$_{d1,i}$, C$_{d2,j}$, p)) + cardinality(entities) \label{lst:line:cpcost}
              \EndIf
          \EndFor
\EndFor
\State \Return CPs, count
\EndFunction
\end{algorithmic}
\end{algorithm*}

Algorithm~\ref{algo:computeCP} describes in more detail how to compute FCPs only based on the pre-computed statistics \textit{local\_objects}$_{d1}$ and \textit{local\_subjects}$_{d2}$ (\textsf{newFunction(0)} returns a new function with default value 0). 
First, all common entities in \textit{local\_objects}$_{d1}$ and \textit{local\_subjects}$_{d2}$ are identified in line~\ref{lst:line:overlap}. These common entities represent links between CSs C$_{d1,i}$ and C$_{d2,j}$ via  property $p$ and are captured by a FCP (lines~\ref{lst:line:newcp}-\ref{lst:line:cpcost}). 

\mbox{
\begin{lstlisting}[basicstyle=\scriptsize\sffamily,caption={Find LMDB movies that are also DBpedia movies},label=queryA,language=sparql,numbers=none,frame=none,columns=fixed,extendedchars=true,breaklines=true,showstringspaces=false]
SELECT ?film ?movie WHERE {
 ?film dbo:budget ?budget .
 ?film dbo:director ?director .
 ?movie owl:sameAs ?film .
 ?movie lmdb:sequel ?seq
}
\end{lstlisting}
} \\
FCPs can be used for cardinality estimation and join ordering using the same principles as described in Section~\ref{sec:cp}. 
Consider a federation consisting of DBpedia (160,061 CSs) and LMDB (8,466 CSs) with 22,592 FCPs and query in Listing~\ref{queryA}. 
We can use Formula~\ref{f:estimationMultCP} with the FCPs connecting LMDB to DBpedia via the owl:sameAs property to estimate the result cardinality: 171. 
This is close to the real cardinality (293). 
\remove{If instead of using Algorithm~\ref{algo:computeCP} to compute FCPs between DBpedia and LMDB we use SPARQL queries, it will require more than a billion expensive queries and 40 years if each query were answered in a second.}

\subsection{Reducing the Sizes of Entity Descriptions}
\label{sec:reduceSizes}

\begin{figure*}[htb]
	\centering
	\includegraphics[width=0.8\textwidth]{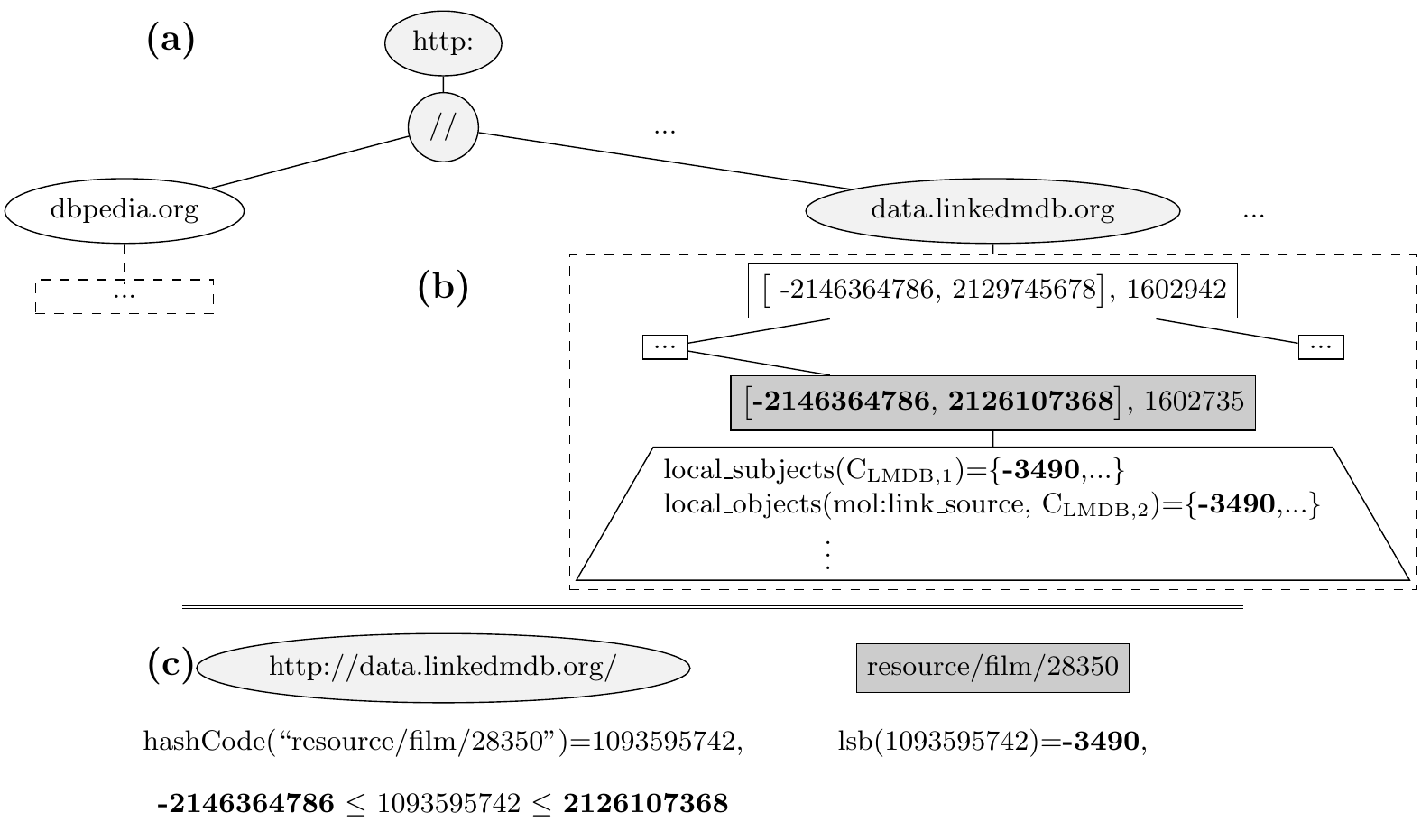}
	\caption{Reduced Entity Descriptions for LMDB in Fig.~\ref{fig:federatedStatistics}. The tree factorizes common prefixes in the top level (in the ellipses) and summarizes the suffixes in the middle (in the rectangles) and bottom (in the trapezium) levels}
	\label{fig:reducedStats}
\end{figure*}

As the entity descriptions (\textit{local\_subjects}$_{d}$ and \textit{local\_objects}$_{d}$) introduced above are often very expensive to compute, maintain, and exchange, we propose a technique to reduce their sizes. 
We organize the entity descriptions in a tree structure that summarizes the entities used as subject or object in any of the dataset's triples.  
Inspired by~\cite{DBLP:conf/www/HarthHKPSU10,DBLP:conf/edbt/PrasserKK12,DBLP:conf/esws/SaleemN14}, we factorize common prefixes, transform suffixes into integers, and summarize sets of integers in buckets, 
i.e., a set synopsis consisting of minimum value (mn), maximum value (mx), \textit{\big[mn, mx\big]}, number of elements, \textit{num}, and their set of two least significant bytes (\textit{lsb}). \textit{lsb}(i) is computed as $i$ \textbf{mod} $2^{16}$ and is included to improve the synopsis' accuracy.

The tree structure is organized in three levels. The top level summarizes the prefixes of entity IRIs occurring as subjects and objects in the dataset. Suffixes are mapped to integers using a hash function, these integers are summarized in the middle and bottom levels. The middle level includes buckets where parent nodes subsume the synopsis of their children (containment relationship between parent and child ranges and summation between parent and child \textit{num}) and aids in efficiently accessing the bottom level.
The bottom level (leaves) stores (in \textit{local\_subjects} and \textit{local\_objects}) only the integer's \textit{lsb} to reduce the storage space while improving the synopsis' accuracy.

In Fig.~\ref{fig:reducedStats} we present a fragment of the reduced descriptions for LMDB. The reduced descriptions  include all the entities that are subject or object in the dataset's triples. In particular, it includes the entity with IRI \textit{http://data.linkedmdb.org/resource/film/28350} (Fig.~\ref{fig:reducedStats}(c)). 
This IRI prefix identifies the subtree that summarizes the entity (light gray ellipses in Fig.~\ref{fig:reducedStats}(a)), while the hash code of its suffix (resource/film/28350), 1093595742, is used to identify the leaf that includes its lsb (-3490), i.e., with 1093595742 between its minimum and maximum values (gray rectangle in Fig.~\ref{fig:reducedStats}(b)). Its lsb is in \textit{local\_subjects}(C$_{\text{LMDB},1}$) and \textit{local\_objects}(mol:link\_source, C$_{\text{LMDB},2}$) in the identified leaf (trapezium in Fig.~\ref{fig:reducedStats}(b)). 
This tree structure exhibits size reduction and eases the computation of FCPs by allowing to discard large portions of the descriptions contrary to descriptions in  Fig.~\ref{fig:federatedStatistics}(b), where all the \textit{local\_subjects} and \textit{local\_objects} need to be pair-wise tested for overlap.

Computation costs are greatly reduced by pruning large portions of the tree and comparing only a few pairs of leaves, the ones that have common prefixes and overlapping  representation of the suffixes. An important feature of these summaries is that entities present in more than one dataset are always detected.
 
These trees are  considerably lighter than the entity descriptions discussed in Section~\ref{sec:fedStats}, but they might reduce accuracy. 
For FedBench's DBpedia 3.5.1 subset, a dataset with 43,126,772 triples that occupies 6.1GB, the \textit{local\_subjects} and \textit{local\_objects} occupy 1.37GB and the tree occupies only 68MB\footnote{Implementation based on Java's HashSet and HashMap was used to measure their sizes.}. They have compression ratios of 4.45 and 91.86, respectively. Regarding the quality, the tree summary allows for computing all the FCSs and FCPs.

To reduce the resources used by the tree, we have reduced the number of CSs as suggested in~\cite{DBLP:conf/edbt/Gubichev014,DBLP:conf/icde/NeumannM11} to 10,000.
Only the CSs that are shared by the largest number of entities are kept, and the others are removed and merged into the remaining CSs if possible. 
For instance, by selecting from the remaining CSs that include all the properties of the removed CS, the one with the smallest number of properties and combining their   \textit{count} and \textit{ocurrences}, 
or by splitting the removed CS into two disjoint property sets that can be merged with other CSs. 
This may reduce the accuracy of the query cardinality estimation, but it allows to bound the resources used to store and access these statistics. 

\remove{Some SPARQL query optimization approaches such as~\cite{DBLP:conf/semweb/SaleemNPDH13} use lighter summaries (MIPs). However, in our context, MIPs could miss some federated CPs. For instance, MIPs allow to compute only 13\% of the federated CPs for DBpedia. While our summaries allow to compute 100\%.}

Entity summaries can be kept up-to-date in two ways. For datasets that are rarely updated, the subtree representing the entities with the prefix affected by the updates, e.g., Fig.~\ref{fig:reducedStats}(b) in our example, can be re-computed. For datasets that are often updated, leaves should support removal of entities, this can easily be done by storing the multiplicity of each least significant byte  so they are removed only if all the entities with that least significant byte have been removed from the dataset.

\newbox\mybox
\begin{lrbox}{\mybox}
\begin{lstlisting}[basicstyle=\scriptsize\sffamily,language=sparql]
SELECT DISTINCT * WHERE {
 ?f dbo:budget ?b .                (tp1)
 ?f dbo:director ?d .              (tp2)
 ?m owl:sameAs ?f .                (tp3)
 ?m lmdb:sequel ?s .               (tp4) 
 ?d dbo:birthDate ?bd .            (tp5)
 ?d dbo:activeYearsStartYear ?sy   (tp6)
}
\end{lstlisting}
\end{lrbox}

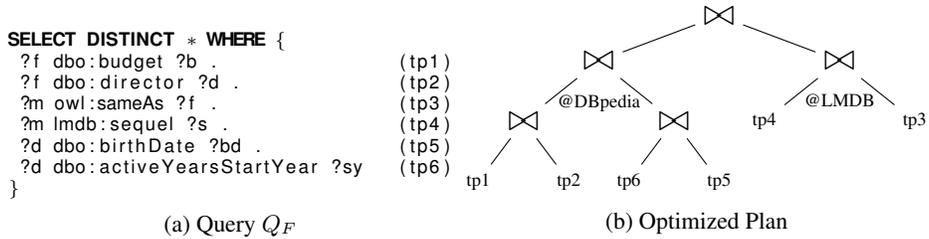
\begin{figure*}[htb]
\raisebox{1cm}{\subfloat[Query $Q_F$]{\label{fig:queryD}\usebox\mybox}}%\hspace*{-1cm}
\subfloat[Optimized Plan]{\label{fig:planD}
\begin{tikzpicture}[font=\scriptsize,level 1/.style={sibling distance=32mm, level distance=6mm}, level 2/.style={sibling distance=20mm, level distance=8mm},level 3/.style={sibling distance=12mm,level distance=8mm},level 4/.style={sibling distance=20mm,level distance=14mm},level 5/.style={sibling distance=13mm,level distance=14mm}]
\node{\Large $\bowtie$}
child{ node(sqA){\Large $\bowtie$}
          child{ node(sqAA){\Large $\bowtie$}
	          child{ node(tp2) {tp1}}
	          child{ node(tp1) {tp2}}
          }
          child{ node(sqBAa){\Large $\bowtie$}
		child{ node(tp4a) {tp6}}
		child{ node(tp3a) {tp5}}
	}
}
child{ node(sqBAb){\Large $\bowtie$}
		child{ node(tp4b) {tp4}}
		child{ node(tp3b) {tp3}}
	}
;
\node[node distance=0.11cm,below=of sqA] {@DBpedia};
\node[node distance=0.09cm,below=of sqBAb] {@LMDB};
\end{tikzpicture} }
\caption{Query $Q_F$ and its Optimized Plan}
\label{fig:planDP}
\vspace{-2mm}
\end{figure*}

\subsection{Optimizing Federated Queries}
\label{sec:optimization}

Query optimization in \opti can logically be divided into the following steps:
\begin{enumerate*}
 \item preprocessing and source selection, 
 \item join ordering, and
 \item query decomposition.
\end{enumerate*}
Arbitrary queries can be handled incrementally by optimizing its subqueries. In the following, we address the optimization of queries with bound predicates, \opti relies on existing optimizers to handle other queries.

\subsubsection{Preprocessing and source selection}

We first parse the query and identify its star-shaped subqueries. Then, properties in each star-shaped subquery are used to identify relevant CSs and sources.  For example, the subquery composed by tp3 and tp4 in Fig.~\ref{fig:planDP}(a) has relevant CSs that include both owl:sameAs and  movie:sequel. In the FedBench federation described in Table~\ref{tab:numberCSCP}, these CSs are only part of LMDB. Therefore, LMDB is the only relevant source for this subquery.
Afterwards, we use CPs/FCPs to identify relevant sources for the links between the star-shaped subqueries. 

\subsubsection{Join ordering}
\begin{figure}
	\centering
	\includegraphics[width=1\textwidth]{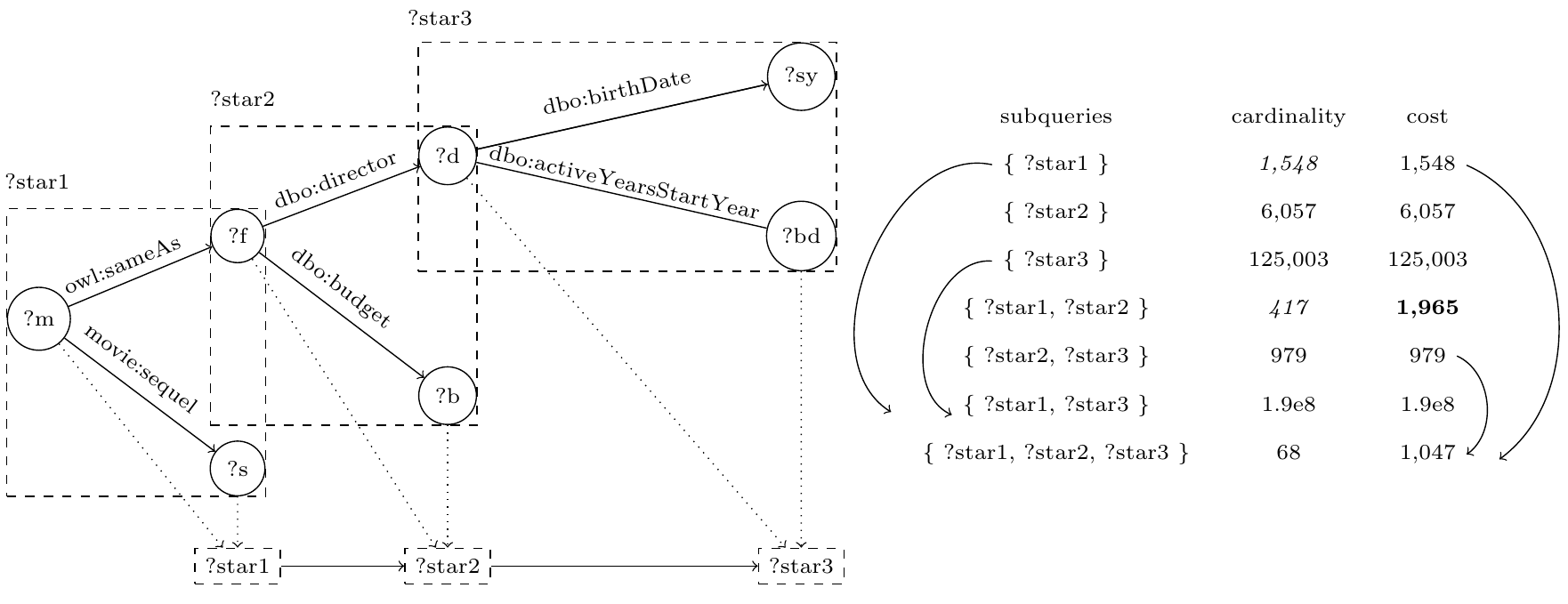}
	\caption{Example query optimization}
	\label{fig:opti}
	\vspace{-2mm}
\end{figure}

Once we have identified the set of relevant sources, we can estimate cardinalities of subqueries and find the best join ordering. We first optimize the order of joins and triple patterns within each star-shaped subquery using CS statistics (\textit{count(C$_i$)} and \textit{occurrences(p,C$_i$)}) as explained in Section~\ref{sec:cp}. 
Afterwards, as described in Section~\ref{sec:cp}, each subquery is treated as a meta-node and 
we estimate cardinalities of the joins between these meta-nodes using the formulas presented in Section~\ref{sec:cp} to estimate subquery costs and apply DP.  
For $Q_F$ (Fig.~\ref{fig:planDP}(a)), three star-shaped subqueries are identified and treated as meta-nodes to estimate the cardinalities of their joins (Fig.~\ref{fig:opti}, left). 
Fig.~\ref{fig:opti} (right) shows the estimated cardinality and cost of the subqueries, solid arrows indicate which smaller subqueries were combined by the DP algorithm to form larger subqueries.  
As the number of subqueries is usually considerably lower than the number of triple patterns, applying DP becomes affordable. 

In our current implementation, the cost function is solely defined on the cardinalities of intermediate results and how many results need to be transferred from endpoints during execution. This favors query plans with selective subqueries. 
For instance, the cost of the join between meta-nodes ?star1 and ?star2 (\textbf{1,965}) includes the result size (\textit{417}) and the sum of all transferred intermediate results (\textit{1,548}). 
This cost function assumes that all endpoints have the same characteristics. We can easily extend this cost function by additional parameters that can be fine-tuned to represent the characteristics of each endpoint individually, e.g., communication delays, response times, etc. 

\subsubsection{Query decomposition}
Finally, we optimize the SPARQL queries that are actually sent to the endpoints and try to minimize their number. For instance, we combine all triple patterns and logical subqueries to a particular endpoint into
 a single SPARQL query to a particular endpoint whenever possible. For instance, meta-nodes ?star2 and ?star3 in Fig.~\ref{fig:opti} are combined into one subquery (Fig.~\ref{fig:planDP}(b)) and evaluated by the DBpedia endpoint.  

\section{Evaluation}
\label{sec:experimental-results}

\begin{table*}[htb]
	\vspace*{-2ex}
	\caption{FedBench~\cite{DBLP:conf/semweb/SchmidtGHLST11} dataset statistics: number of distinct triples (\#DT), predicates (\#P), CSs (\#CS), and CPs (\#CP); computation time in seconds of \opti, HiBISCuS, and voiD statistics}
	\label{tab:numberCSCP}
	\begin{center}
		\begin{scriptsize}
			\begin{tabular}{|c|c|c|c|c|c|c|c|c|}
				\hline
				Dataset & \#DT & \#P & \# CS & \#CP & \#FCP & \opti & HiBISCuS & voiD \\
				\hline
				ChEBI& 4,772,706           & 28 & 978 & 9,958                  & 19,360 & 82.91 & 96.02 & 73.89\\
				KEGG & 1,090,830           & 21 & 67 & 239                     & 13,822 & 30.15 & 95.23 & 12.84\\
				Drugbank & 517,023        & 119 & 3,419 & 12,589                  & 103,070 & 1,299.9 & 76.4 & 6.98\\
				DBpedia subset & 42,855,253      & 1,063 &10,000 & 1,069,431 & 6,583 & 2,739 & 770.48 & 1,465.36\\
				Geonames & 107,949,927 & 26 &673 & 7,707                  &322,672 & 1,885.97 & 609.52 & 39,694.07\\
				Jamendo & 1,049,647        & 26 &42 & 190                       &1,259  & 31.25 & 99.17 & 14.66\\
				SWDF & 103,595               & 118 &547 & 6,713                    &17,557 & 7.27 & 69.21 & 2.03\\
				LMDB & 6,147,916            & 222 &8,466 & 94,188                 &359,340& 947.16 & 317.21 & 355.45\\        
				NYTimes & 335,119          & 36 &47 & 158&3.96 & 10.01 & 72.56 & 4.22\\
				\hline
				Federated & \multicolumn{5}{c|}{} & 620.35 & \multicolumn{2}{c}{} \\				
				\cline{1-1}
				\cline{7-7}
				\cline{7-9}
				Total & \multicolumn{5}{c|}{} & 7,654.27 & 2,205.8& 41,629.5\\												
				\cline{1-1}
				\cline{7-9}
			\end{tabular}    
		\end{scriptsize}
	\end{center}
	\vspace*{-3ex}
\end{table*}

In this section, we present the results of our experimental study that compares our approach, \opti, with state-of-the-art federated query engines: HiBISCuS (FedX-HiBISCuS, cold and warm cache)~\cite{DBLP:conf/esws/SaleemN14}, SemaGrow~\cite{DBLP:conf/i-semantics/CharalambidisTK15}, FedX (cold and warm cache)~\cite{DBLP:conf/semweb/SchwarteHHSS11}, and SPLENDID~\cite{DBLP:conf/semweb/GorlitzS11}.
Full implementations, statistics, and results are available at \url{\texttt{https://github.com/gmontoya/federatedOptimizer}}.

\noindent \textbf{Datasets and queries:} We use the real datasets and queries proposed in the FedBench benchmark~\cite{DBLP:conf/semweb/SchmidtGHLST11}. Queries are divided into three groups Linked Data (LD1-LD11), Cross Domain (CD1-CD7), and Life Science (LS1-LS7). They have 2-7 triple patterns and star and hybrid shapes. They  have between 1 and 9,054 answers. Basic statistics about the datasets are listed in Table~\ref{tab:numberCSCP}. 
We ran each query ten times and report the averages over the last nine runs. Standard deviation is included as error bars on the plots.

\noindent \textbf{Implementation:} \opti is implemented in Java using the Jena library to parse and transform queries into queries with SPARQL 1.1 service clauses. 
Our implementation uses the FedX 3.1 framework with deactivated native optimization to execute \opti's query plans. 

\noindent \textbf{Hardware configuration:} 
For our experiments we used virtual machines (VMs). A VM   
using up to 4GB of RAM to run the federated query engine and 
nine VMs with 2 processors, 8GB of RAM and CPU 2294.250 MHz 
to host Virtuoso endpoints with the datasets 
described in Table~\ref{tab:numberCSCP} (one dataset and endpoint per VM). 

\noindent \textbf{Statistics computation:}
As DBpedia has a very high number of CSs (160,061), we reduced them to 10,000 by merging (as suggested in~\cite{DBLP:conf/edbt/Gubichev014,DBLP:conf/icde/NeumannM11} and explained in Section~\ref{sec:reduceSizes}) without significant losses in the quality of estimations. 
Details on creation times of statistics are listed in Table~\ref{tab:numberCSCP}. \opti's statistics can be more expensive to compute for datasets with more than 3,419 CSs and cheaper than HiBISCuS's for datasets with less than 67 CSs. In total, \opti's statistics are computed five times faster than voiD's.

\remove{
\noindent {\bf Statistical analysis:}
The Wilcoxon signed rank test~\cite{wilcoxon1992individual} for paired non-uniform data is used to study the significance of the improvements on performance obtained when \opti query optimization is used\footnote{The Wilcoxon signed rank test was computed using the R project (\url{http://www.r-project.org/}).}. It computes a probability value (p-value) that can be used to decide whether an approach significantly outpferforms another one; if the p-value is lower than 0.05, we can conclude that \opti significantly outperforms the other approaches.}

\noindent \textbf{Evaluation metrics:}
\begin{enumerate*}
\item {\it Optimization time (OT):} is the elapsed time since the query is issued until the optimized query plan is produced, 
\item {\it number of selected sources (NSS):} is the number of sources that have been selected to answer a query, 
\item {\it number of subqueries (NSQ):} is the number of subqueries that are included in the query plan, 
\item {\it execution time (ET):} is the time elapsed since the evaluation of the query plan starts until the complete answer is produced 
(with a timeout of 1,800 seconds), 
\item {\it number of transferred tuples (NTT):} is the number of tuples transferred from all the endpoints to the query engine during query evaluation.
\end{enumerate*}

\noindent \textbf{Result completeness:}
All approaches produce the complete result set for non-timed out queries, except SPLENDID for query LS7. 

\subsection{Experimental Results}

\begin{figure*}[htb]
	\centering
	\includegraphics[width=0.92\textwidth]{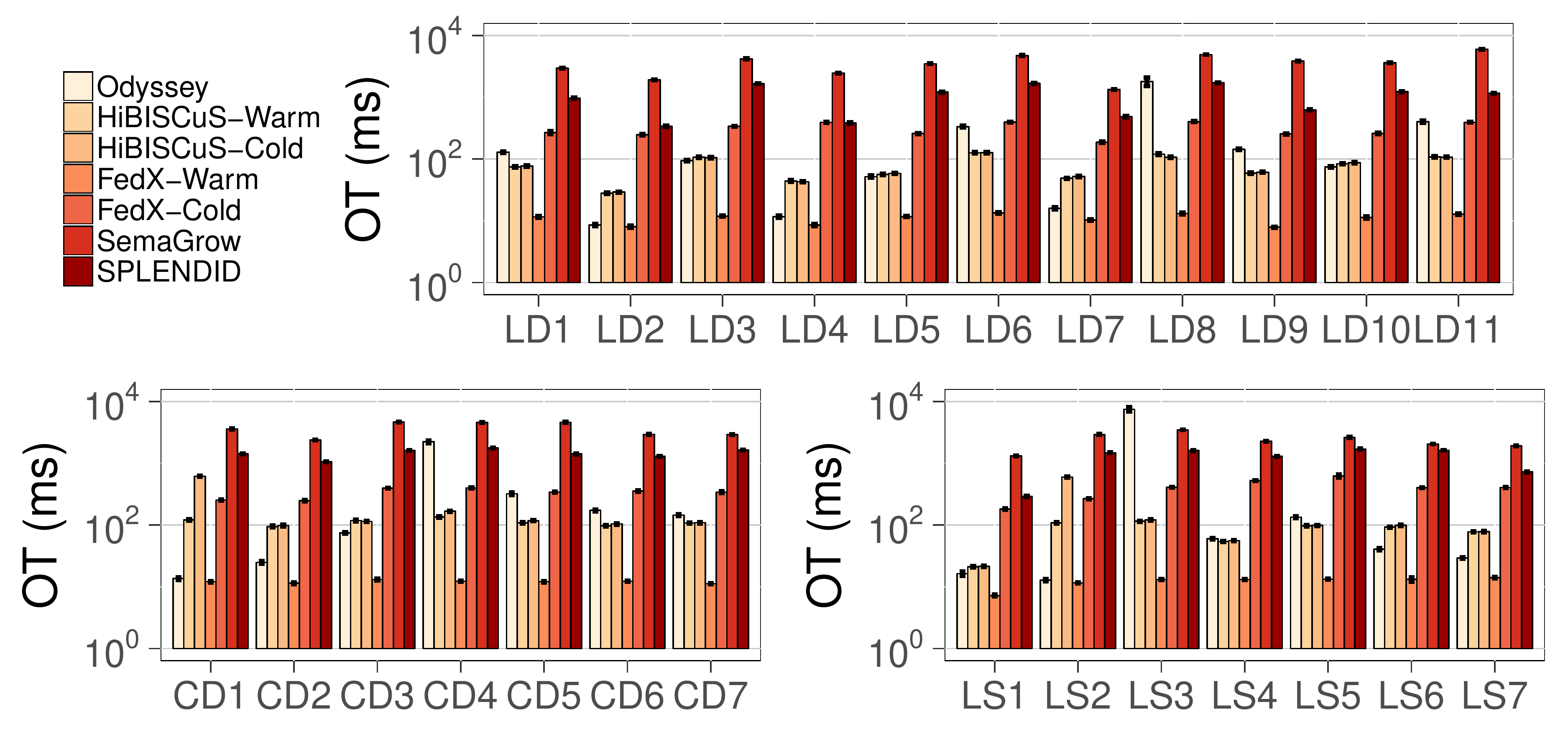}
	\caption{Optimization Time in ms (OT, log scale). CD1 and LS2 have variable predicates and \opti relies on FedX to find plan.}
	\label{fig:PT}
	\vspace{-2mm}
\end{figure*}

\subsubsection{Optimization time}
Fig.~\ref{fig:PT} shows the optimization time (OT) for the studied approaches. 
Because of the detailed statistics and dynamic programming, one might expect \opti to suffer from a considerable overhead in OT. As our experimental results show, however, \opti's query planner is competitive to most other approaches with a slight advantage for FedX-Warm as this system has cached information about the query relevant sources. 
For instance, \opti is up to 69 times faster (SemaGrow) than other approaches on average. 
\remove{P-values lower than 0.05 confirm that \opti optimization is significantly faster than SemaGrow  and SPLENDID.} 

\begin{figure*}[htb]
	\centering
	\includegraphics[width=0.92\textwidth]{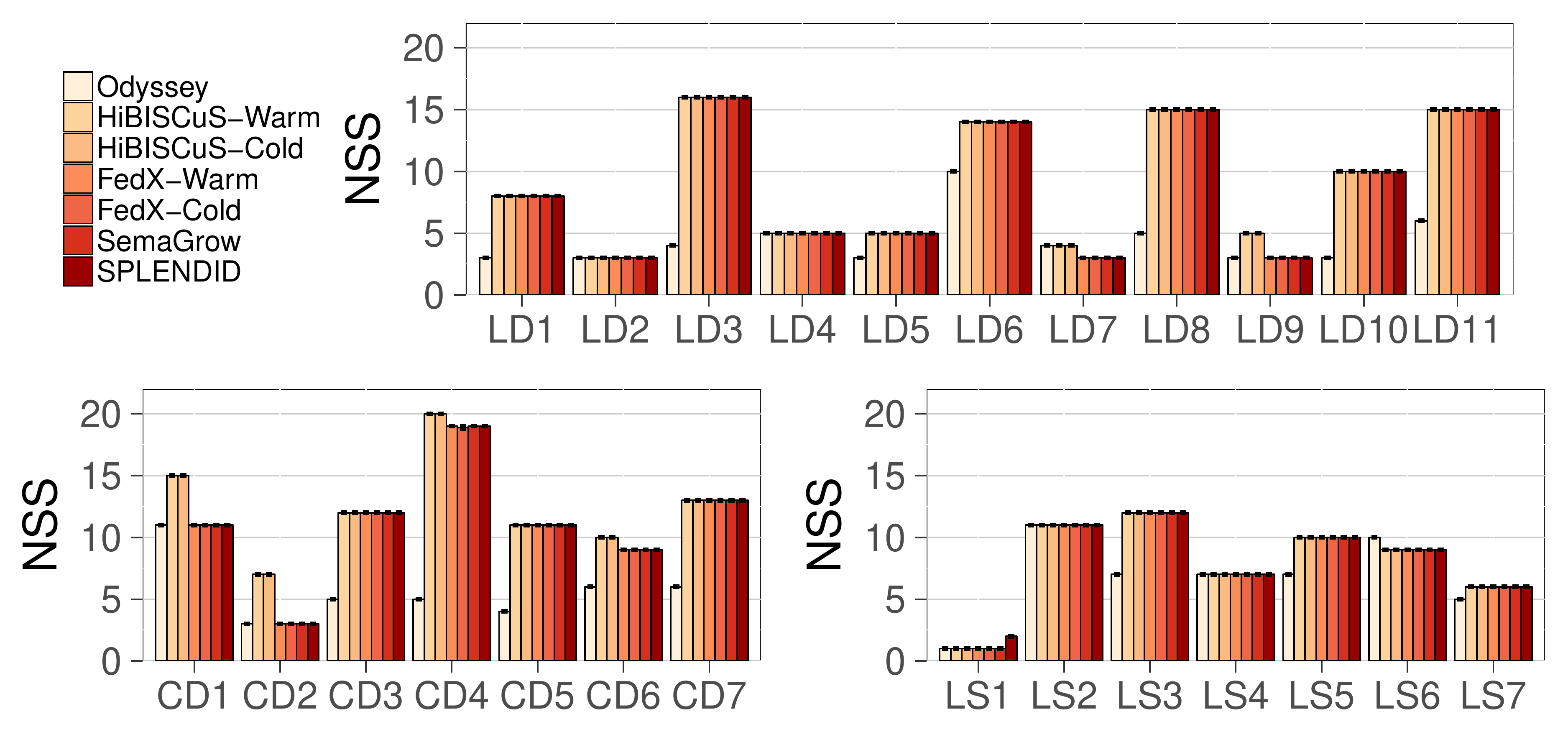}
	\caption{Number of Selected Sources (NSS)}
	\label{fig:NSS}
	\vspace{-2mm}
\end{figure*}

\subsubsection{Number of selected sources}
As Fig.~\ref{fig:NSS} shows, \opti selects only a small number of relevant sources; for instance, at least 1.81 times less (FedX-Cold/Warm and SemaGrow) and up to 1.93 times less (HiBISCuS-Cold/Warm) on average. 
\remove{P-values lower than 0.05 confirm that \opti selects significantly fewer sources than all the other approaches.}
For some queries, e.g., LS4, existing approaches already select the optimal number of sources.  
For LD7, \opti selects a larger number of sources than the optimum because our approach does not perform ASK queries during execution to prune irrelevant sources. Sometimes \opti overestimates the set of relevant sources -- but on the other hand it never misses any relevant sources.
For LS1, most approaches select just one (10$^0$) source because there is only one dataset that has triples with the predicate in the query. 

\begin{figure*}[htb]
	\centering
	\includegraphics[width=0.92\textwidth]{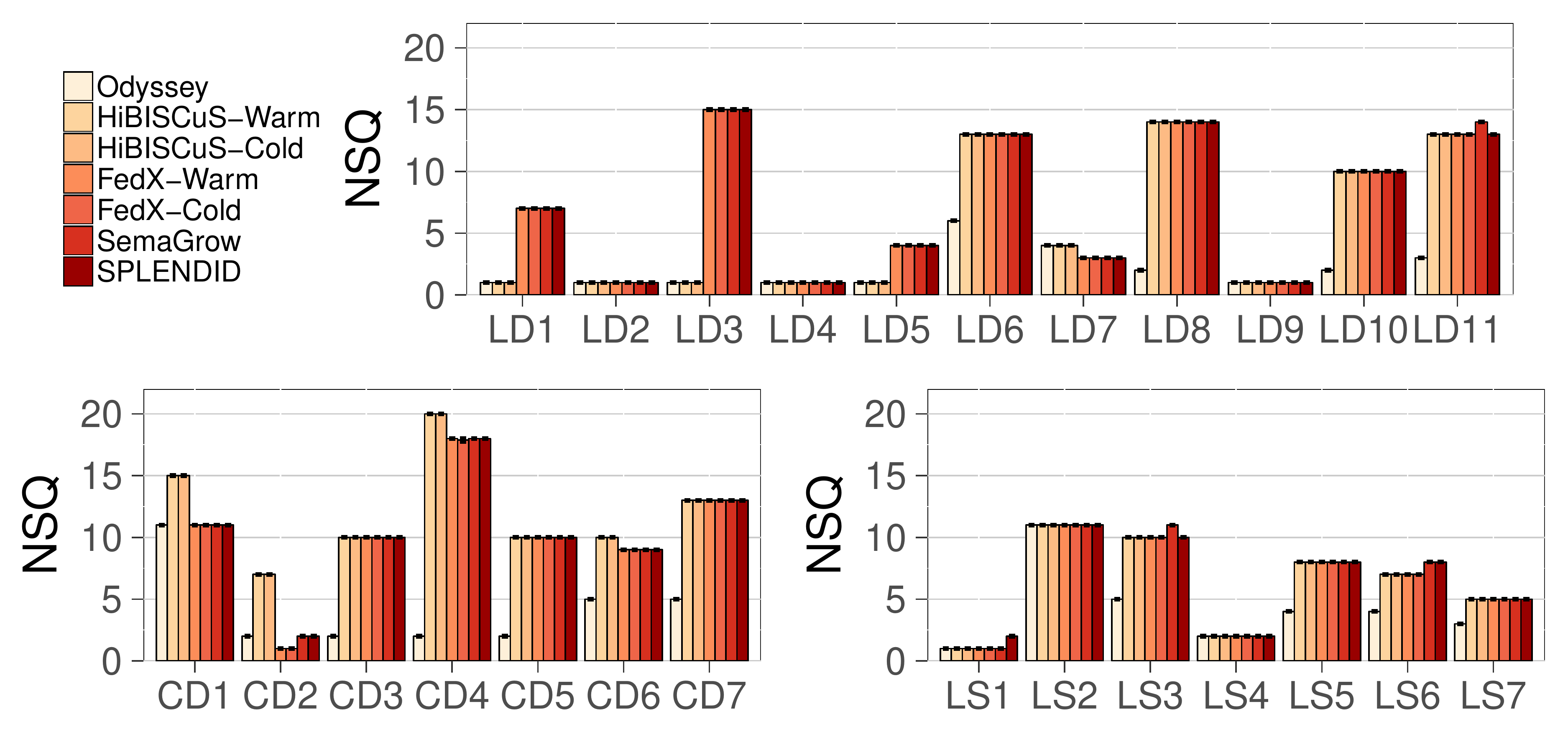}
	\caption{Number of Subqueries  (NSQ)}
	\label{fig:NSQ}
	\vspace{-2mm}
\end{figure*}

\subsubsection{Number of subqueries}
As Fig.~\ref{fig:NSQ} shows, \opti uses considerably fewer subqueries than other approaches, at least 2.62 times less (HiBISCuS-Cold/Warm) and up to 3.41 times less (SPLENDID) on average. 
\remove{P-values lower than 0.05 confirm that \opti decomposes the queries in significantly less subqueries than all the other approaches.}
The fact that Odyssey always produces the correct and complete answers confirms that \opti correctly identifies and exploits cases for which it is advantageous to combine subqueries. \opti's reduction of the number of relevant sources has a positive impact on the number of subqueries (NSQ), \opti's pruning of non-relevant sources allows for combining triple patterns into subqueries without affecting the result completeness. 
Some queries, like LD2, LD4, and LD9, include triple patterns that can be evaluated by a unique endpoint of the federation and existing approaches already decompose the query into the optimal NSQ. 
Only for LD7, FedX-Cold/Warm, SPLENDID, and SemaGrow decompose the query into fewer subqueries than \opti, this is because they use ASK queries to assess a source's relevance.  \opti could be enhanced with this strategy.

\begin{figure*}[htb]
	\centering
	\includegraphics[width=0.92\textwidth]{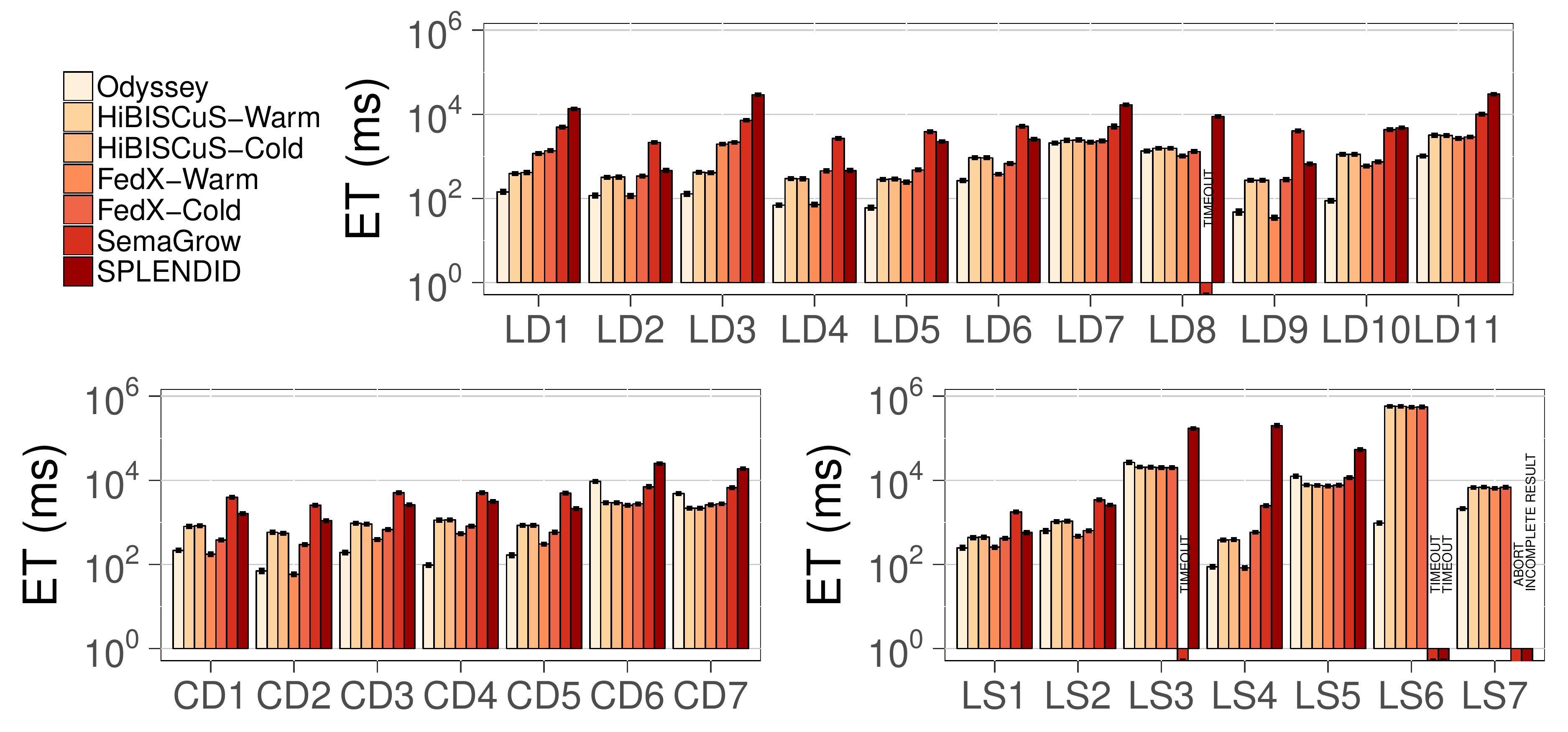}
	\caption{Execution Time in ms (ET, log scale)}
	\label{fig:ET}
	\vspace{-2mm}
\end{figure*}

\subsubsection{Execution time}
Some approaches failed to answer all queries before the timeout (1,800s): SPLENDID (2 queries) and SemaGrow (4 queries).  
Even when considering only those queries that completed before the timeout, \opti is on average 126.26 times faster than SPLENDID and 28.30 times faster than SemaGrow. 
Fig.~\ref{fig:ET} shows the execution times (ET) for the studied approaches. \opti is on average at least 25.46 times faster (FedX-Warm). 
\remove{P-values lower than 0.05 confirm that \opti is significantly faster than \hibiscus-Cold/Warm \ and FedX-Cold.}
Only for a few queries \opti is (slightly) slower than other approaches, e.g., LS3. As for the other metrics, \opti's ET can be improved if ASK queries were used during query execution to further reduce the relevant sources similarly as it is done by other approaches. For five of the queries,  \opti is one of the fastest approaches and for 11 queries, \opti is the fastest approach.
\opti's achieved reductions on the NSS and NSQ have a positive impact on the ET; as fewer endpoints  are queried fewer times, \opti produces results faster than most approaches in most cases.

\begin{figure*}[htb]
    \centering
    \includegraphics[width=0.92\textwidth]{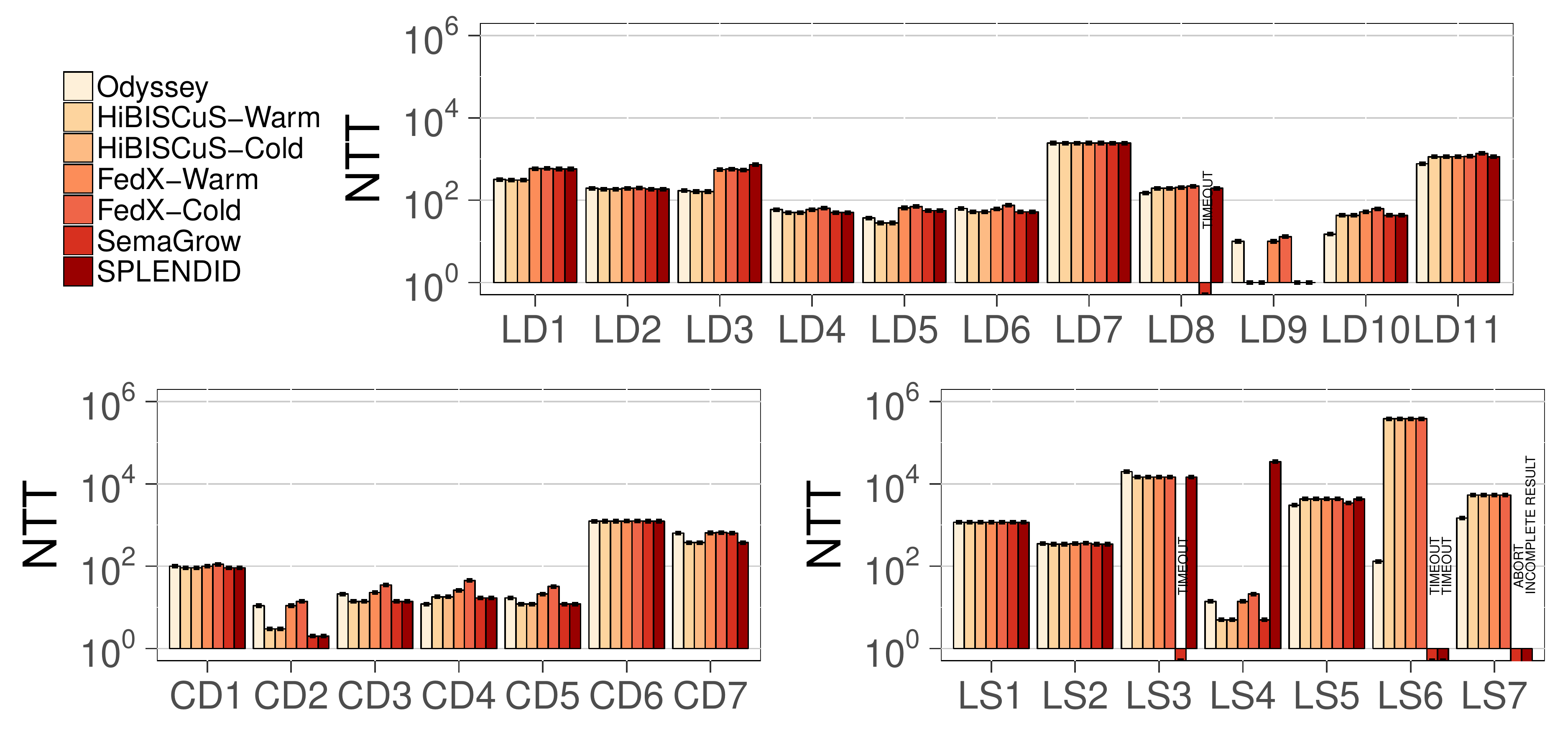}
    \caption{Number of Transferred Tuples  (NTT, log scale, 10$^0$=1)}
    \label{fig:NTT}
    \vspace{-2mm}
\end{figure*}

\subsubsection{Number of transferred tuples}
Fig.~\ref{fig:NTT} shows the number of transferred tuples (NTT) for the studied approaches. 
\opti transfers fewer tuples than other approaches. 
Even when considering only those queries that completed before the timeout, \opti transfers on average 1.15 times fewer tuples faster than SemaGrow and 108.4 times fewer tuples than SPLENDID. 
For the approaches that completed all the queries, \opti transfers at least 117.55 fewer tuples (HiBISCuS-Cold/Warm) on average. 
\remove{P-values lower than 0.05 confirms that \opti transfers significantly less tuples than FedX-Cold/Warm.}
Most approaches are competitive in terms of NTT. The largest difference is observed for LS6, where \opti clearly outperforms the other approaches transferring 500 times fewer tuples.  
In contrast to other approaches, \opti not only reduces the number of requests sent to the endpoints but also avoids non-selective subqueries, which significantly reduces network traffic and the local query load at the endpoints.

\begin{figure*}[htb]
	\centering
	\includegraphics[width=0.92\textwidth]{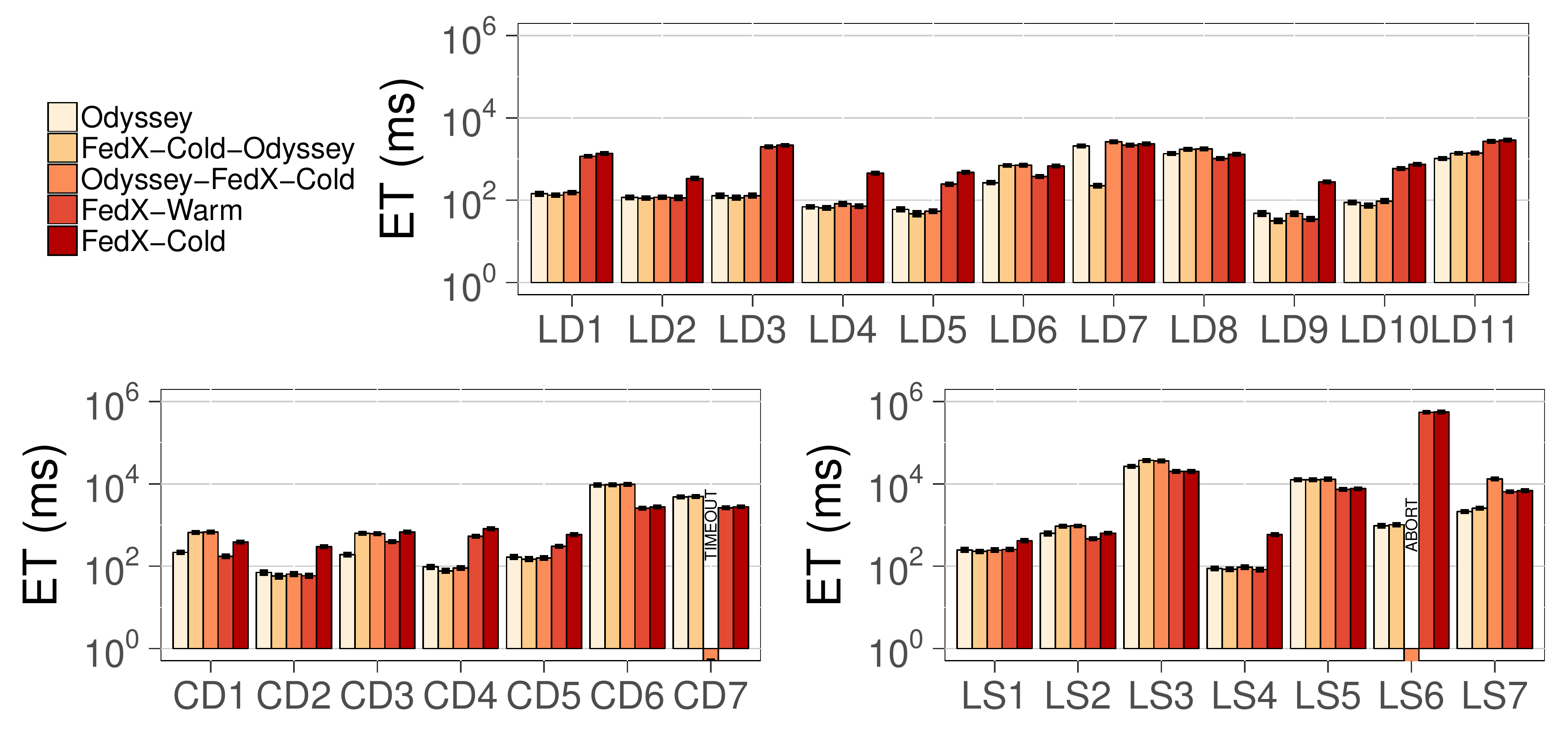}
	\caption{Execution Time in ms  (ET, log scale)}
	\label{fig:ETH}
	\vspace{-2mm}
\end{figure*}

\subsection{Combining \opti with Existing Optimizers}

We have also integrated \opti techniques directly into the FedX optimizer and obtained: 
\begin{itemize}
 \item \opti-FedX-Cold, which relies on CSs and CPs to select sources and decomposes the query but uses FedX join ordering.
 \item FedX-Cold-\opti, which relies on the FedX optimizer for source selection but uses \opti for query decomposition and join ordering.
\end{itemize}
Fig.~\ref{fig:ETH} compares the execution times (ET) of these two implementations with \opti, FedX-Cold, and FedX-Warm. 
In most cases the combined approaches are considerably faster than native FedX. 
In a few cases, however, their ET can increase considerably. 
In these cases, queries include a highly selective subquery with one triple pattern and using FedX's heuristic to execute subqueries with more than one triple pattern first leads to plans that are more expensive than others.
On average, the combined approaches are 26.86 and 3.99 times faster than FedX-Cold. 
\remove{P-values lower than 0.05 confirm that FedX-Cold-\opti is significantly faster than FedX-Cold  and FedX-Warm.}

For query LD7, \opti and FedX-Cold/Warm exhibit similar ETs whereas FedX-Cold-\opti is considerably faster. For this query it happens that the advantages of both \opti and FedX coincide, i.e., we can take advantage of the good join ordering by \opti but also of the additional pruning based on ASK queries by FedX. 
\\\\
\noindent Even if \opti's OT can be higher in comparison to existing approaches,
\opti produces better plans composed of fewer subqueries and fewer selected sources per triple pattern without compromising result  completeness. Benefits of these features have been evidenced with significantly faster ETs and less transferred data from endpoints to the federated query engine.

\section{Conclusion}
\label{sec:conclusion}

In this paper, we have presented \opti, an approach for optimizing federated SPARQL queries based on statistics.  
These statistics detail information about the data provided by remote endpoints as well as the links between them. This enables more accurate cost estimations, query optimization, and selection of relevant sources. 
Our extensive experimental evaluation shows that \opti produces query execution plans that are better in terms of data transfer and execution time than state-of-the-art optimizers.
In our future work, we plan to further improve \opti by considering in which situations exactly it is worthwhile to use additional aspects of other optimizers, such as ASK queries and associated statistics. Another interesting direction of future work is to further reduce the computation time and sizes of the entity descriptions and provide efficient strategies to update the descriptions and statistics.

\subsection*{Acknowledgments}
This research was partially funded by the Danish Council for Independent Research (DFF) under grant agreement no. DFF-4093-00301. The final publication is available at Springer via \url{http://dx.doi.org/10.1007/978-3-319-68288-4\_28}.

\bibliographystyle{abbrv}
\bibliography{references}

\end{document}